# Cell lineage tracing using nuclease barcoding

Stephanie Tzouanas Schmidt[1], Stephanie M. Zimmerman[2,#a], Jianbin Wang[1,#b], Stuart K. Kim[2,3], Stephen R. Quake[1,4,5*]

[1] Department of Bioengineering, Stanford University, Stanford, California, United States of America

[2] Department of Genetics, Stanford University, Stanford, California, United States of America

[3] Department of Developmental Biology, Stanford University, Stanford, California, United States of America

[4] Department of Applied Physics, Stanford University, Stanford, California, United States of America

[5] Howard Hughes Medical Institute, Stanford University, Stanford, California, United States of America

[#a] Department of Genome Sciences and Medicine, University of Washington, Seattle, Washington, United States of America

[#b] School of Life Sciences, Tsinghua University, Beijing, China

* Corresponding author
Email: quake@stanford.edu

Short title: Nuclease-based cell lineage tracing

## *Abstract*

Lineage tracing, the determination and mapping of progeny arising from single cells, is an important approach enabling the elucidation of mechanisms underlying diverse biological processes ranging from development to disease. We developed a dynamic sequence-based barcode for lineage tracing and have demonstrated its performance in *C. elegans*, a model organism whose lineage tree is well established. The strategy we use creates lineage trees based upon the introduction of specific mutations into cells and the propagation of these mutations to daughter cells at each cell division. We present an experimental proof of concept along with a corresponding simulation and analytical model for deeper understanding of the coding capacity of the system. By introducing mutations in a predictable manner using CRISPR/Cas9, our technology will enable more complete investigations of cellular processes.

## *Introduction*

From the beginning of the study of biology, scientists have been interested in lineage tracing, or the determination and mapping of progeny arising from single cells [1]. Numerous approaches have been realized and utilized over time, with the earliest attempts involving careful sectioning and staging of embryos [2] as well as direct observation of developing animals [3, 4]. As direct observation is not implementable for all systems, labeling of cells has been used to study a wide variety of questions in an assortment of experimental systems and organisms, first by dyes [5, 6] and radioactive tracers [7] and later by genetic markers [8] and genetic recombination [9,10].

While existing methods can identify the descendants of a cell, inferring the relationships between descendants remains to be adequately addressed, and all techniques thus far present drawbacks that render high-throughput and high-resolution lineage tracing a difficult undertaking. Intended for the study of tissues in which genetic lineage tracing is challenging and incomplete, the approach we present involves lineage tree construction based on the propagation of mutations arising in individual cells to their daughter cells.

Though prior work has explored lineage analysis based upon naturally arising mutations [11], we have engineered a strategy by which cell lineage trees can be constructed based upon the introduction of specific mutations into cells and the propagation of these mutations to daughter cells at each cell division. With the advent of clustered regularly interspaced short palindromic repeat nucleases (CRISPR) as a genome editing tool, tractable and specific targeting of genomic sites has been made possible [12, 13]. By creating mutations in a predictable manner using CRISPR/Cas9, this technology will enable more complete investigations of cell differentiation in processes ranging from development to cancer emergence.

## *Results*

### CRISPR/Cas9-mediated strategy for updateable sequence barcodes

To engineer dynamic, sequence-based barcodes, we used CRISPR/Cas9 genome editing to target ten specific sites within the gene encoding enhanced green fluorescent protein (EGFP) in the EG6173 strain of *C. elegans*. Our strategy relies upon non-homologous end joining (NHEJ) to repair the double-stranded breaks caused by CRISPR/Cas9 and to introduce insertion-deletion mutations (indels) at the specified sites contained within the genome of a given cell. After such a site has been mutated, the indel introduced makes it unavailable for further alteration by CRISPR/Cas9 and serves as an identifiable marker of the cell and its daughters (Fig. 1).

Given the diversity of indels generated by NHEJ, the accumulation of indels at the targeted sites over the course of several generations of cell division creates a barcode within a cell's DNA that allows for the construction of lineage trees based upon the indels shared between the barcodes of different cells (Fig. 1A). That the sites designated for CRISPR/Cas9 editing occur within a region of 500 base pairs allows for the barcode to be easily amplified and read using paired-end sequencing. However, it is important to note that if CRISPR/Cas9 acts upon a second site before a given site can be repaired, the processes of NHEJ can result in the removal of the sequence between the two sites, potentially omitting useful information (Fig. 1B). In spite of this, our simulations show that the dynamic DNA barcode approach could enable the tracing of cell lineages with potentially greater resolution and specificity than existing methods and without the need for continuous monitoring of the system of interest.

## Mathematical prediction of system behavior

To characterize and better understand the CRISPR/Cas9-enabled barcoding technique, we simulated sequence barcodes and the resulting lineage trees resulting from its implementation in *C. elegans*, a nematode and model organism whose cell lineage tree was established in the late 1970s by means of real-time microscopic observation [4]. We represented CRISPR/Cas9 activity as a Poisson process and followed the known cell division pattern from *C. elegans* for the initial four divisions during which major cell lineages are specified. Barcodes were generated to represent two cases, in which NHEJ did and did not result in dropouts, the removal of sequence between targeted sites.

For both cases and for a range of expected values of the Poisson process, 100 sets of simulated barcodes were generated and corresponding lineage trees constructed. Modulating the expected value of the Poisson process enabled us to explore the effects of the CRISPR/Cas9 cutting rate, which we defined as the number of indels introduced per time. To compare the agreement of each simulated tree with the known lineage tree, we calculated the cophenetic correlation between the two. As cophenetic distance describes the intergroup similarity of the cluster in which a pair of observations is placed within a tree structure, the correlation between the resulting cophenetic distance matrices of full tree structures provides a suitable metric for the evaluation of concordance between two trees [14].

With this in mind, we computed the cophenetic correlation of each simulated tree with the known tree (Fig. 2). We found that the correlation improved with increasing numbers of indels introduced during each generation, but even with approximately one indel introduced per cell division, corresponding to a Poisson expected value of 0.05 cuts per minute, the lineages constructed agreed well with the reference lineage. As expected, the presence of dropouts reduced the agreement with the known tree, but despite this, for lower Poisson expected values, the difference in performance between barcode sets with and without dropouts was not statistically significant. However, it is important to note that even with the cutting rate approaching 8 cuts per cell division, the best correlations achieved were 0.99 without dropout and only 0.75 with dropouts. This suggests that in any practical situation the stochastic generation of lineage barcodes will not enable perfect lineage reconstruction unless the scheme is modified to eliminate dropouts.

## Barcoding *C. elegans*

We tested our proposed barcoding strategy by using it to label cells in *C. elegans*. Ten sites within the sequence encoding EGFP, codon optimized for *C. elegans,* were selected for targeting by CRISPR/Cas9, thus collectively serving as a sequence barcode. The stochastic creation of indels by CRISPR/Cas9 and NHEJ over the course of a worm's development was expected to generate a diverse collection of resulting sequence barcodes corresponding to different cells within the organism. Due to the sequential addition of indels during development, the indels observed in the barcodes obtained from the adult worm could then be used to infer the relationships between the cells from which the barcodes came, since once a cell's barcode is modified to include a particular indel, the barcodes of all subsequent daughter cells must also carry that indel.

Ribonucleoprotein complexes (RNPs) of Cas9 and single guide RNA (sgRNA) corresponding to each of the ten sites were injected into the gonad of worms from the EGFP-expressing EG6173 strain so that the RNPs would be encapsulated within the forming eggs. After 48 hours post injection we screened the F1 progeny of the parent animals for absence of EGFP expression, since successful targeting by CRISPR/Cas9 of the selected sites within the EGFP gene would lead to the introduction of indels, thus disrupting the fluorescent protein (Fig. 3). Progeny displaying the desired phenotype were dissected in order to remove their intestine, since this organ is known to derive from the descendants of a single cell, the E blastomere [4, 15]. The isolated intestine and remainder of the body were lysed and sequenced separately, and the EGFP gene containing the sites targeted was inspected for the presence or absence of indel mutations.

The indels contained within the sequence barcodes enabled the determination of lineage relationships between cells. Since indels are transmitted from parent to daughter cells, barcodes displaying shared indels can be considered to constitute a sublineage. We grouped barcodes by such common indels and used the frequency of indels observed across unique barcodes to infer the order of introduction of indels and thus, to construct lineage trees, a subset of which is shown in Fig. 4. We found that barcodes from the intestine displayed an assortment of indels not observed in the rest of the body, suggesting a distinct cell type in agreement with the known lineage, since indels created after the formation of the E blastomere should be restricted to appearing in either the intestine or the body. Moreover, in both the sample derived from the intestine and that from the rest of the body, sets of barcodes contained shared indels, indicating descent from a common progenitor cell, such as the P1 or EMS blastomeres that occur prior to the formation of the E blastomere. The presence of barcodes such as the leftmost two shown that contain no indels or just one indel in the adult worm reveals that the rate of CRISPR/Cas9 activity needs to be further tuned to record cell divisions over the entirety of development.

To assess the performance of the proposed strategy in correctly capturing the separation of the intestinal lineage from the rest of the body, the correspondence of the generated barcodes to tissues of origin was determined using the k-nearest neighbors algorithm [16]. By comparing the predicted tissue of origin to the actual tissue of origin for each barcode, we were able to calculate two instructive metrics of information retrieval: recall, the ratio of correctly identified intestinal barcodes to the total number of actual intestinal barcodes, and precision, the ratio of correctly identified intestinal barcodes to the total number of putatively identified intestinal barcodes (Fig. 5) [17]. This analysis was carried out using the full collection of barcodes derived from the sequenced intestine and body of the worm shown in Fig. 4, and the results were compared to those generated by randomizing the dataset's barcode assignments over 100 trials. The experimental results show that the precision of the actual data is 0.86 compared to 0.32 for the randomized case, while the recall is 0.40 compared to 0.03 for randomization. These results underscore the ability of updateable sequence barcodes to capture lineage relationships using approaches independent of tree construction.

Information content of nuclease barcoding

To examine the diversity of indels generated using CRISPR/Cas9 and to calculate the information content of the described barcoding method, the indels observed across the ten targeted sites in eight sequenced *C. elegans* samples were compared (Fig. 6). The distribution of positions contained within indels relative to the position of the CRISPR/Cas9-introduced doubled-stranded break (DSB) for cases in which dropouts did not occur can be described using a Gaussian curve (Fig. 6a). The Shannon entropy calculated from this curve is 4.42 bits [18].

The theoretical upper bound of information provided by CRISPR/Cas9 barcoding was determined using a simple analytical model in which the probability of position occupancy by an indel conditional upon the indel's length was used. Making use of the parameters describing the distribution of experimental indel lengths and assuming a uniform distribution for the utilization of positions themselves, we found that the Shannon entropy of an individual cut site within such a sequence barcode could encode 6.86 bits of information at most. To provide a more realistic estimate, we weighted the occupancy of particular positions by their likelihood of inclusion in the indel itself, giving a Shannon entropy of 6.61 bits.

The difference between the theoretical values obtained and the empirical result can be explained in part by the bias observed in the creation of indels (Fig. 6b). By describing the location of the DSB with respect to the center of the subsequent indel normalized to the length of the indel, it is evident that the indels resulting from CRISPR/Cas9 activity tend to lie to the left of the specified DSB, reducing the number of possible states and amount of information able to be encoded through barcoding. This point can be further illustrated by employing a modified version of the aforementioned analytical model. By using a delta function instead to describe the occupancy of positions around the DSB to determine a lower bound on performance, we calculated a Shannon entropy of 3.80 bits for a single cut site, which is still a considerable improvement over the effectively binary information supplied through current recombination- and color-based lineage tracing techniques.

## *Discussion*

We report a novel technology enabling dynamic sequence barcoding for lineage tracing. *C. elegans*, an organism with a completely known lineage, served as the model system for our work, and CRISPR/Cas9 was used to perform genome editing. The diverse indels introduced at specified sites over the course of the animal's lifespan allowed for the creation of unique and updateable barcodes contained within a cell's genome. The indels shared between barcodes allowed for the inference of parent-daughter relationships between cells consistent with the organism's well-established lineage tree. Despite the occurrence of what we have termed dropouts, the system generated barcodes that distinctly labeled lineages known to be separate from one another. The mathematical modeling work we performed supports the experimental lineage tracing results obtained.

Lineage tracing remains an important technique in the study of biology. While color-based and inducible systems are widely utilized in the study of lineages, current approaches are limited in the resolution provided, the ability to combine lineage information with other system read-outs, and the need for real-time observation [19]. We were motivated by the great potential for data storage and manipulation within DNA and thus developed our dynamic, sequence-based barcoding technology. Previous work has exploited the natural occurrence of mutations in microsatellite loci [11] or mitochondrial DNA [20] to trace lineages. A recent report also describes a CRISPR/Cas9-enabled barcoding strategy for lineage tracing, which was implemented in a cell line and in zebrafish, although it did not analyze the coding capacity of the system [21]. Our work demonstrates the potential of such nuclease-based barcoding methods by illustrating the agreement between the known lineage of *C. elegans* and that obtained using dynamic sequence barcodes, delves into the challenge presented by sequence dropouts through a simulation-based approach, and underscores the technique's promise by establishing its information content.

With this in mind, controllably introducing mutations in a desired genomic region and at rates higher than background offers the ability to track parent-daughter relationships with even greater clarity and without the need for whole-genome amplification. Though earlier approaches, e.g. zinc finger nucleases [22] and TALENS [23], have been used to edit genomes, the flexibility and specificity afforded by CRISPR/Cas9 made it attractive for the strategy we have described. Certainly, future work to develop control of CRISPR/Cas9 activity or to explore other methods of introducing mutations into DNA could improve the performance of such a system towards achieving the predicted upper limit of information content of 6.61 bits. With that much information in each site and just ten sites as we used in our experiments, one could create approximately $7.9 \times 10^{19}$ unique barcodes, more than enough to track all of the $3.7 \times 10^{13}$ cells in the human body. Taken together, we present proof of concept for the tracking of cells by means of dynamic sequence-based barcodes, which we are confident will enable unprecedented and more complete studies of cell lineage.

## *Materials and Methods*

### Mathematical Modeling

The activity of CRISPR/Cas9 can be represented as a Poisson process, which can be represented by the following:

$$P(n \text{ indels in a given time}) = \frac{\lambda^n e^{-n}}{n!}$$

Where $\lambda$ is the average number of indels created per minute and $n$ is the number of indels considered. A Python script was written and implemented to update theoretical sequence barcodes corresponding to those expected in *C. elegans* [3] based upon the introduction of indels whose occurrence followed a Poisson distribution within cells following the known division times and patterns. Across a range of values used for $\lambda$, similar to an indel introduced every other generation and to eight indels introduced each generation, barcodes were generated with and without dropouts, the removal of genetic material between cut sites. Cophenetic correlation was then used to compare the theoretical lineage trees determined from the modeled barcodes to the known lineage tree [14].

## *C. elegans* Maintenance

*C. elegans* expressing enhanced green fluorescent protein (EGFP) were obtained from the University of Minnesota's Caenorhabditis Genetics Center and served as the wild-type strain for these experiments. The nematodes were maintained per the established protocols [24].

## Ribonucleoprotein Complex Preparation and Injection into *C. elegans*

Ten EGFP-specific sgRNAs were designed and transcribed using the MEGAshortscript T7 Transcription Kit from Ambion; purified Cas9 was purchased from PNA Bio. The pooled sgRNAs and Cas9 protein were mixed at a 1:2 mass ratio and incubated for 10 minutes at 37 °C [25]. The sgRNA-Cas9 complexes were injected into the gonads of wild-type young-adult P0 animals, and the resulting F1 progeny from the following 48 hours were screened for knockdown of EGFP.

## Sequencing and Analysis

F1 progeny were screened for EGFP knockdown. Selected worms were lysed and sequenced whole, while others were dissected to remove the intestine. Each intestine, corresponding body, or whole body was lysed separately per previously described methods [26]. The region encoding EGFP in each lysate was PCR-amplified and underwent paired-end sequencing using the Illumina MiSeq system. Low-quality and low-abundance reads were removed, and only reads joined by FLASH were used [27]. A custom Python pipeline was used to align the obtained reads to the reference sequence over each selected cut site via Smith-Waterman alignment [28] and identify indels introduced by the activity of the sgRNA-Cas9 complexes.

Only indels, all of which included or were within 3 base pairs of the DSB, were considered in order to mitigate potential amplification and sequencing errors within the barcodes. Each sequence was then expressed as a vector in which the presence or absence of each indel observed across the collection of indels was recorded using ones and zeros, respectively, to create binary barcodes. Maximum parsimony was implemented in R to reconstruct the relationships between unique binary barcodes and create lineage tree representations [29, 30]. To evaluate the accuracy and precision of the barcodes in correctly identifying different tissue types, the k-nearest neighbors algorithm was executed in R for k = 3 [16]. Randomized assignments of the experimental dataset were generated and the results compared to those from the actual assignments.

## Calculation of Information Content

The distributions of lengths and positions occupied by indels were determined, and from the position distribution, the experimental Shannon entropy of the system was calculated using the known expression:

$$H = p \times \log p$$

To determine the upper bound of information able to be encoded within a single cut site, the following was considered:

1. Assuming that the distribution of indel lengths, *l*, is Gaussian with mean, $\mu_l$, and standard deviation, $\sigma_l$, as determined empirically, we write:

$$p(l) = \Phi\left(\frac{l - \mu_l}{\sigma_l}\right)\Big]_{l-0.5}^{l+0.5}$$

2. We can express the probability of a particular position, *d*, being occupied in terms of the corresponding indel's length as follows:

   a. Uniform Distribution

$$p(d|l) = \frac{1}{2l}$$

   b. Weighted Distribution

$$p(d|l) = \frac{l - |d| + 1}{l + 1}$$

   c. Delta Distribution

$$p(d|l) = \begin{cases} 1, & d \leq l \\ 0, & d > l \end{cases}$$

3. Combining the above two equations and normalizing the probability distribution obtained so that it sums to one,

$$p(d) = \sum_{d=l}^{\infty} p(d|l) \times p(l)$$

4. We can write the Shannon entropy as:

$$H = -\sum_{d=0}^{\infty} p(d) \times \log p(d)$$

These calculations were carried out by numerical integration in Python using the parameters describing the Gaussian curve fit to the empirical distribution of indel lengths.

## *Acknowledgements*

We thank Robert Phillips, John Beausang, and Winston Koh for helpful discussions; Norma Neff and Gary Mantalas for assistance with sequencing; and Ben Passarelli, Derek Croote, and Mark Kowarsky for computing support. *C. elegans* strains were provided by the University of Minnesota's Caenorhabditis Genetics Center, which is funded by NIH Office of Research Infrastructure Programs (P40 OD010440).

S.T.S. is supported by the Fannie and John Hertz Foundation Fellowship, National Science Foundation Graduate Research Fellowship, and Gabilan Stanford Graduate Fellowship. This work was supported by a National Heart, Lung, and Blood Institute (NHLBI) U01HL099995 Progenitor Cell Biology Consortium Grant (S.T.S., J.W., S.R.Q.). S.R.Q. is an investigator of the Howard Hughes Medical Institute.

## *Author Contributions*

Conceived and designed the experiments: STS SMZ JW SKK SRQ. Contributed reagents/materials/analysis tools: SKK SRQ. Performed the experiments: STS SMZ. Analyzed the data: STS SMZ SSK SRQ. Wrote the paper: STS SRQ.

A) NHEJ without dropouts

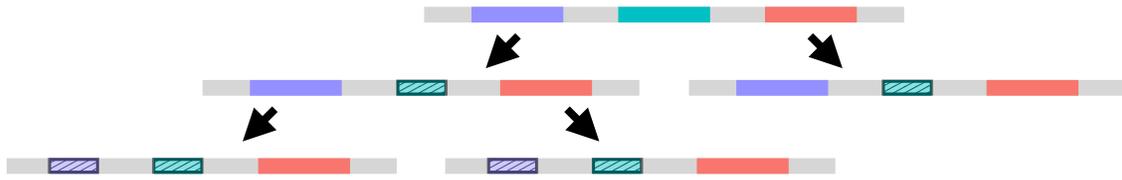

B) NHEJ with dropouts

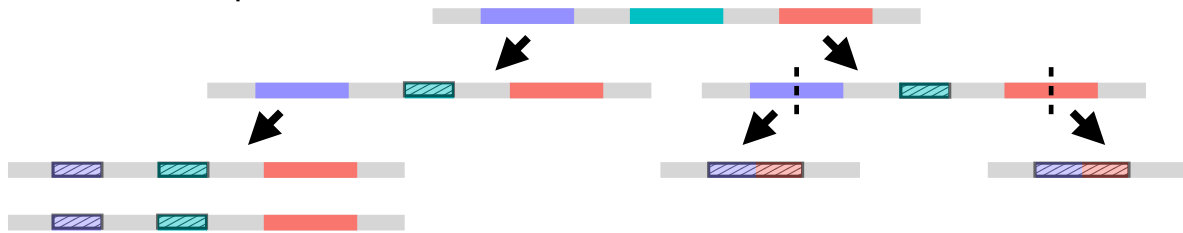

**Fig. 1. Inheritance of introduced mutations enables inference of cell lineages.** Making CRISPR/Cas9 available during development allows for the introduction and transmission of indels in the sequence barcode of cut sites in individual cells. Sample sequence barcodes are represented by grey line segments, and the contained cut sites are represented by each differently colored subsection. Upon targeting by CRISPR/Cas9, the resulting indels are denoted by the shortening of the colored subsections and by the dashed fill pattern. (A) Schematic of representative barcodes in the case where no sequence between cut sites is lost (i.e. no dropouts). (B) Schematic of representative barcodes in the case where dropouts occur, resulting in concomitant loss of information provided by previously created indels.

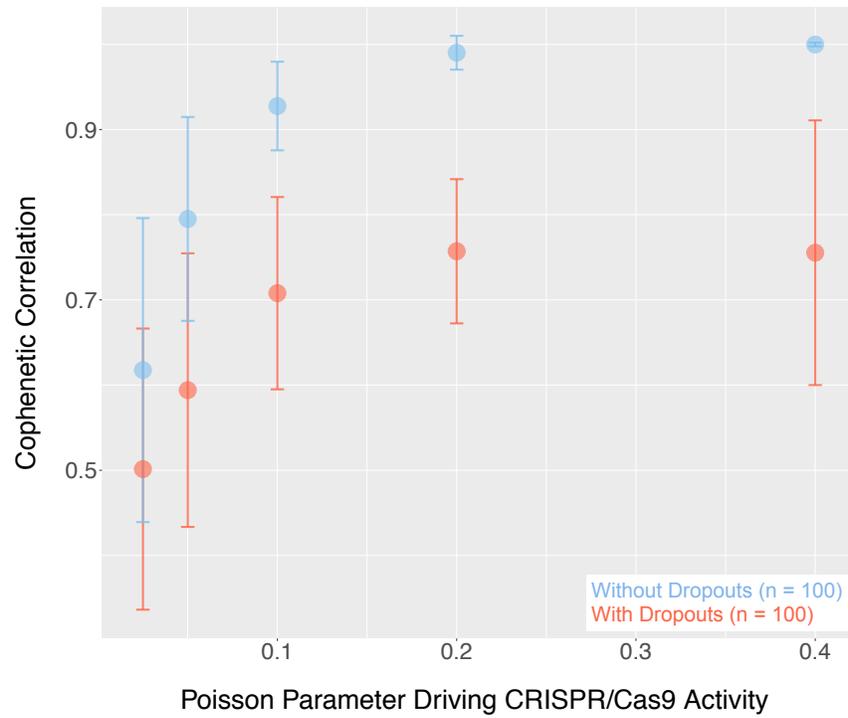

**Fig. 2. Performance overlaps predicted for cases with and without dropouts.** Simulation of sequence barcodes and resulting lineage trees by modeling indel generation as a Poisson process with (blue) and without (red) dropouts enabled comparison with the known *C. elegans* lineage using cophenetic correlation for a range of Poisson expected values (n=100).

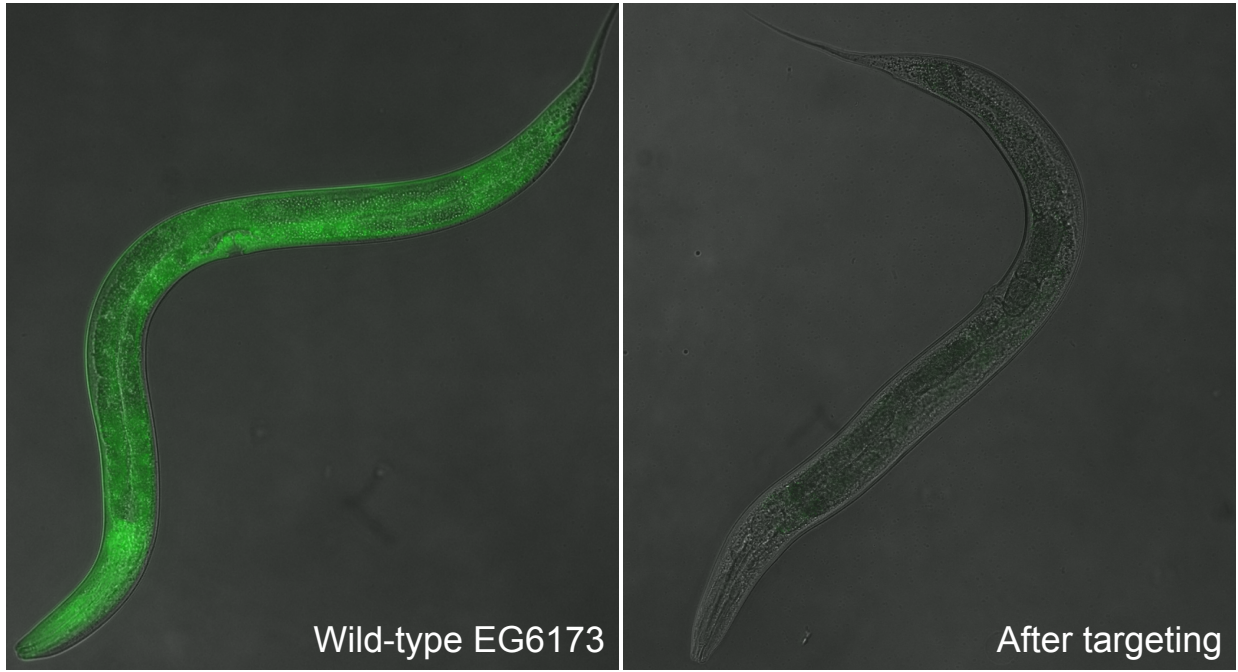

**Fig. 3. Effective CRISPR/Cas9 targeting screenable by phenotype.** Introduction of indels into barcode encoded in EGFP sequence allows for identification of barcoded organisms through disruption of EGFP expression.

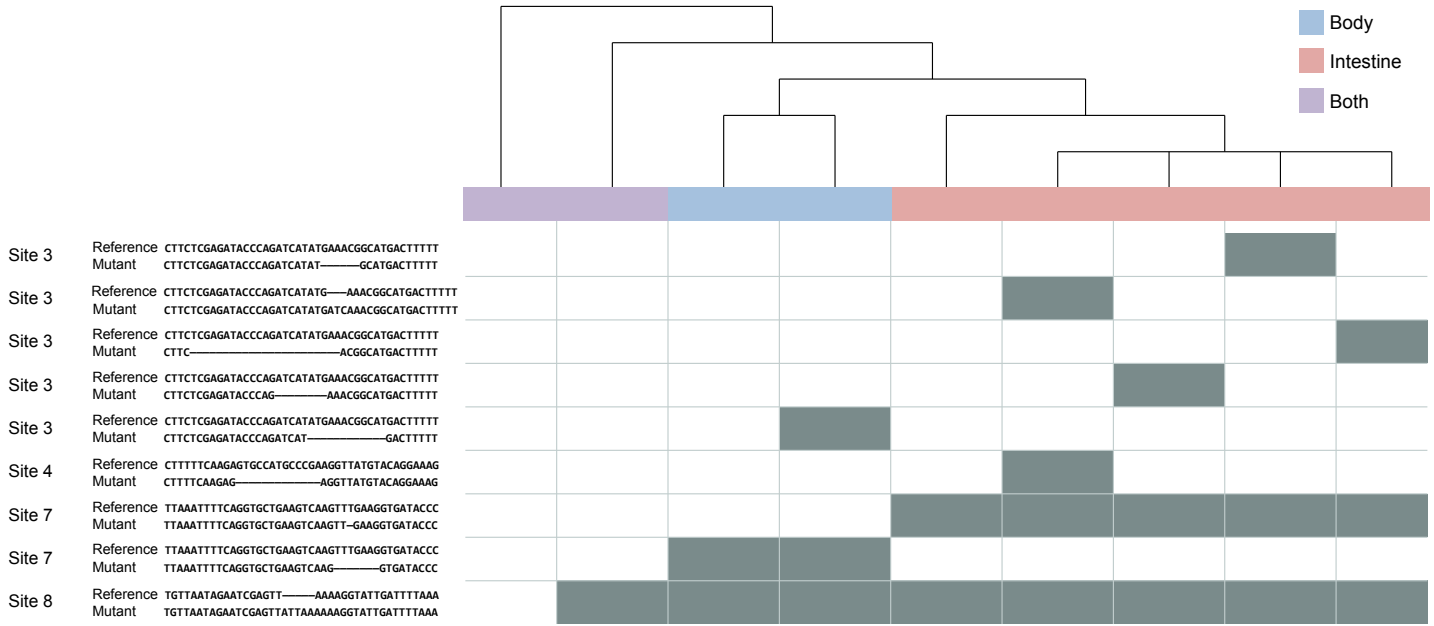

**Fig. 4. Dynamic sequence barcoding identifies distinct cell populations.** The relationships between a subset of unique sequence barcodes derived from the body (blue), intestine (red), and both samples (purple) of a single worm was used to create the following lineage tree. The barcodes corresponding to each member of the tree are presented in the columns of the heatmap, the individual indels are denoted by its rows, and each grey-shaded box indicates the presence of a given indel. The indels are presented in the table to the left of the heatmap as alignments between the reference (top) and observed (bottom) sequences, and the cut site with which each signature is associated is listed in the leftmost column of the table.

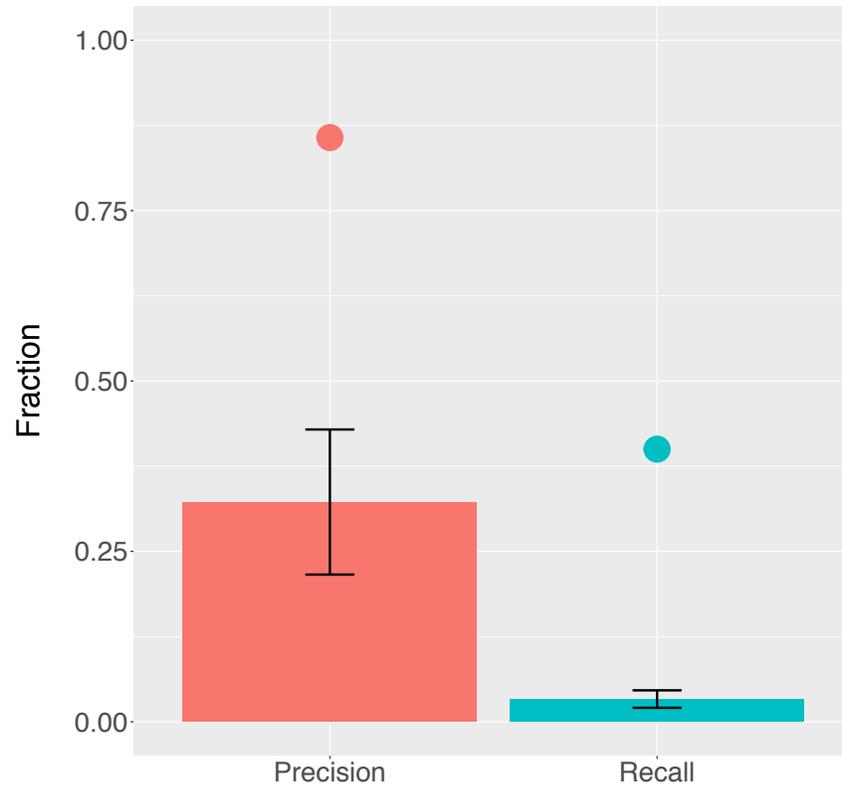

**Fig. 5. Barcoding permits determination of correct tissue of origin.** The precision and recall of tissue identification based upon the full set of barcodes derived from the worm shown in Fig. 4 for which the intestine and body were sequenced separately were calculated using the k-nearest neighbors algorithm. The experimental results, shown by the individual points over the columns, compare favorably with the results obtained by randomizing the dataset's labels (n = 100).

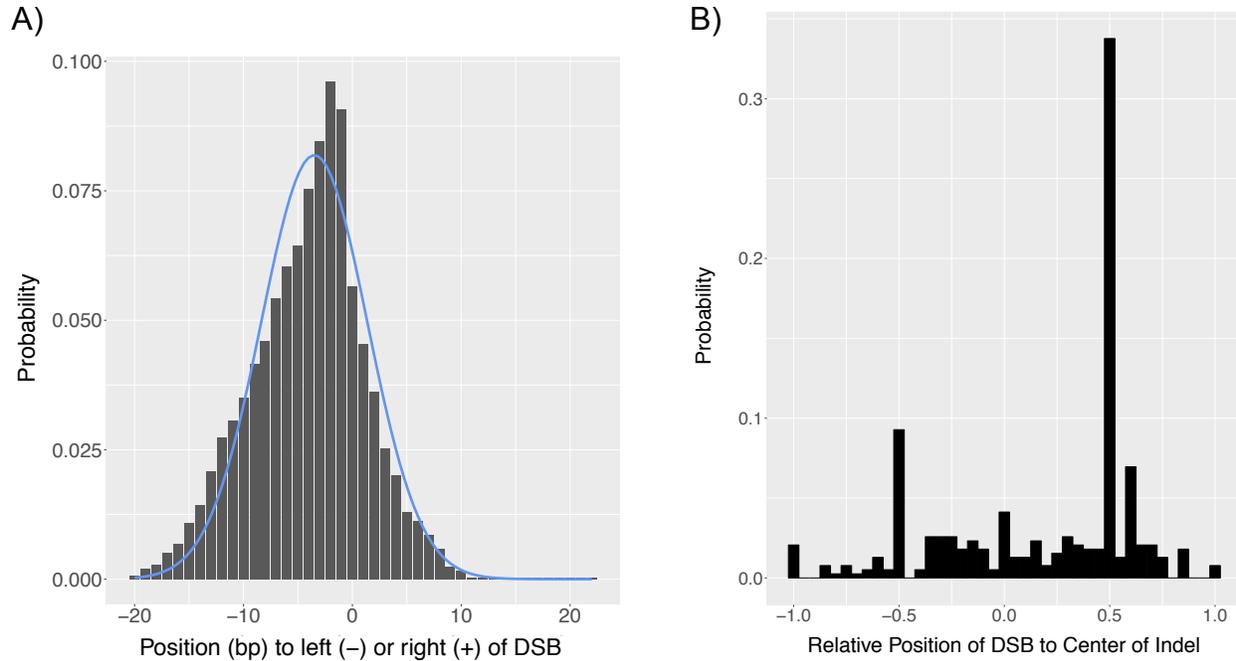

**Fig. 6. Profile of unique indels generated in *C. elegan*s by CRISPR/Cas9 provides insight into information content of proposed barcoding technique.** The unique indels observed in sequenced animals (n = 8) across all ten targeted sites were compared. (A) The probability distribution of which positions flanking the CRISPR/Cas9-created double-stranded break (DSB) were contained in the resulting indels has a Gaussian form. (B) The length-normalized position of the DSB as compared to the center of the indel tends to occur to its right (+) rather than left (-).